\documentclass[10pt,aps,prl,twocolumn,superscriptaddress]{revtex4}

\usepackage[linktocpage=true, colorlinks=true, urlcolor=blue, linkcolor=blue, citecolor=blue]{hyperref}
\usepackage{graphicx}
\usepackage{amsmath,amssymb}
\usepackage{color}
\usepackage{url}
\usepackage{hyperref}
\usepackage{cleveref}
\usepackage{xfrac}
\usepackage{upgreek}
\usepackage{lipsum}
\usepackage{siunitx}

\usepackage[utf8]{inputenc}
\usepackage[T1]{fontenc}

\usepackage{amsmath}
\newcommand*\diff{\mathop{}\!\mathrm{d}}

\graphicspath{{./Images/}}

\begin{document}

\title{Retinal blood flow reversal quantitatively monitored in out-of-plane vessels with laser Doppler holography}

\author{L\'eo Puyo}
\affiliation{Corresponding author: gl.puyo@gmail.com}
\affiliation{Centre Hospitalier National d'Ophtalmologie des Quinze-Vingts, INSERM-DHOS CIC 1423. 28 rue de Charenton, 75012 Paris, France}
\affiliation{Paris Eye Imaging, France}
\affiliation{Institute of Biomedical Optics, University of L\"ubeck. Peter-Monnik-Weg 4, 23562 L\"ubeck, Germany}

\author{Michel Paques}
\affiliation{Centre Hospitalier National d'Ophtalmologie des Quinze-Vingts, INSERM-DHOS CIC 1423. 28 rue de Charenton, 75012 Paris, France}
\affiliation{Paris Eye Imaging, France}
\affiliation{Institut de la Vision. Sorbonne Universit\'e. INSERM, CNRS. 17 Rue Moreau, 75012 Paris, France}

\author{Michael Atlan}
\affiliation{Paris Eye Imaging, France}
\affiliation{Institut Langevin. Centre National de la Recherche Scientifique (CNRS). Paris Sciences \& Lettres (PSL University). \'Ecole Sup\'erieure de Physique et de Chimie Industrielles (ESPCI Paris) - 1 rue Jussieu. 75005 Paris France}

\date{\today}

\begin{abstract}
Laser Doppler holography is a planar blood flow imaging technique recently introduced in ophthalmology to image human retinal and choroidal blood flow non-invasively. Here we present a digital method based on the Doppler spectrum asymmetry that reveals the local direction of blood flow with respect to the optical axis in out-of-plane vessels. This directional information is overlaid on standard grayscale blood flow images to depict flow moving towards the camera in red and flow moving away from the camera in blue, as in ultrasound color Doppler imaging. We show that thanks to the strong contribution of backscattering to the Doppler spectrum in out-of-plane vessels, the local axial direction of blood flow can be revealed with a high temporal resolution, which enables us to evidence pathological blood flow reversals. We also demonstrate the use of optical Doppler spectrograms to quantitatively monitor retinal blood flow reversals.
\end{abstract}

\maketitle

\section*{Introduction}

Doppler ultrasound examinations have become standard practice in clinics to diagnose vascular-related disorders as they offer a good penetration depth, axial sectioning, and a high temporal resolution, while being non-invasive and relatively inexpensive. One of the useful features of Doppler ultrasound is the ability to reveal the axial direction of blood flow. In order to do so, it is necessary to know the sign of the Doppler frequency shift, which in ultrasound is straightforward since the transducer measures the wave frequency. In color Doppler imaging, the local frequency shift is overlaid on anatomical grayscale images to depict flow towards the sensor in red, and flow away from it in blue. In ophthalmology, Doppler ultrasound has however remained limited to the study of large and favorably oriented retrobulbar vessels such as the ophthalmic artery, the central retinal vessels, and posterior ciliary arteries~\cite{Dimitrova2010}. Due to lack of spatial resolution, the inner retinal and choroidal vascular networks still cannot be satisfactorily imaged with Doppler ultrasound.
As for imaging blood flow reversal with Doppler ultrasound, previous reports show arterial retrograde flow in the ophthalmic artery in cases of severe occlusive carotid artery~\cite{Ho1992, Costa1997}, but to our knowledge blood flow reversal in the central retinal artery has not been evidenced with Doppler ultrasound. The spatial resolution of ultrasound waves is insufficient to measure separately the flow of the central retinal artery and vein due to their close proximity in the optic sheath. Consequently, it is assumed that the positive frequencies in the spectrum are arterial flow and negative frequencies are venous flow, so a reversed arterial flow cannot be distinguished from the venous flow. Although retrograde arterial flows are commonly observed in arteries supplying territories of high vascular resistance such as muscles at rest, or in arteries downstream to an arterial stenosis~\cite{Owen2007}, they are not expected to be found in healthy retinas since they are caused by an abnormal downstream or upstream resistance to flow. However, the existence of human retinal arterial retrograde flow has been reported in cases of retinal vein occlusions based on fluorescein or indocyanine-green angiography imaging~\cite{Paques2001, Paques2002, Paques2005}. Prior to that, it was also observed in monkeys with fluorescent vesicle angiography in occlusions of branch retinal vein induced by photocoagulation~\cite{Peyman1998}. Retinal backflow had also already been observed with fluorescein angiography in experiments made on farm pigs eyes in which the intraocular pressure (IOP) was acutely increased~\cite{Dollery1968, Ffytche1974}. It was observed that blood velocity decreases with higher IOP, and that at the level of IOP nearing the no-flow point there is a substantial diastolic backflow in retinal arterioles.
Unfortunately, the observation of blood flow reversal with a fluorescent contrast agent is equivocal as it is unable to objectively quantify the severity of the flow reversal, and challenging because it relies on the observation of the dye front wave. Good fixation is required from the subject and the injection of the dye must be performed rapidly, as otherwise a smooth injection makes the front wave of the dye less observable.

In the past few decades there has been a significant development in optical instruments that non-invasively image and monitor retinal blood flow~\cite{Schmetterer2012, PournarasRiva2013}, the most successful of which are derived from optical coherence tomography (OCT). OCT-angiography provides depth-resolved vascular maps with high spatial resolution, and can be implemented on standard OCT instruments with minimal effort~\cite{Kashani2017, Ang2018}. However, this technique has limited success in measuring blood flow variations. Doppler-OCT is another, albeit more complicated, technique derived from OCT that can measure the local flow velocity in physical units from the phase variations of OCT~\cite{Leitgeb2014}. With Doppler-OCT, blood flow in a given retinal vessel, and eventually total retinal blood flow, can be measured with high temporal resolution and good reproducibility~\cite{Szegedi2020}. Because the sign of the Doppler phase shift is known, Doppler-OCT is sensitive to the flow direction~\cite{Yazdanfar2000}, and is thus able to monitor flow reversal, as demonstrated for example in murine coronary arteries~\cite{Hammer2012}, and in middle cerebral arteries of rat models of ischemic stroke~\cite{ZhuHancock2019}. In ophthalmology, Doppler-OCT has been used to monitor retinal blood flow reversal in Brown Norway rats in which the IOP was acutely increased~\cite{PiJia2019}. However, to our knowledge, retinal blood flow reversal has not yet been reported with Doppler-OCT in the human eye.

Laser speckle contrast imaging (LSCI) derived techniques, and laser Doppler flowmetry (LDF) are two ways of imaging blood flow from the self-interference of light scattered by flowing blood~\cite{Briers1996}. In LSCI, the speckle intensity fluctuations are sampled at low frame rates in order to record spatially/temporally blurred speckle patterns. Blood flow can be quantified in arbitrary units and monitored with high temporal resolution from the blur rate of the speckle figure~\cite{BriersWebster1996, ChengLuo2003, BoasDunn2010}. Due to the accessible frame rates (the camera exposure time is typically in the range of a few milliseconds), some instruments have been commercialized for retinal blood flow imaging~\cite{Sugiyama2010, Debuc2021}. Notably the instrument based on laser speckle flowgraphy (LSFG) has been used in many clinical trials, for example to investigate retinal vascularization in glaucoma~\cite{Shiga2016, Mursch2018}. The biomarkers of interest are usually the local mean blur rate value, and descriptors of the blood flow waveform profile such as the resistivity index.
On the other hand, in LDF the speckle variations are sampled at higher frequencies, faster than the speckle decorrelation time, which arguably offers a better potential to quantify blood flow since the whole Doppler power spectrum density is measured. However, LDF requires a sampling of the intensity fluctuations at much greater frequency than LSCI, which entails receiving more light from the sample. Point-scanning and line-scanning implementations have been demonstrated in the human eye~\cite{RivaLasser1992, Riva2012, Truffer2020, Mujat2019}, with frame rates sometime up to 100 kHz. A full-field version of LDF has also been proposed to image blood flow beneath human skin~\cite{SerovLasser2005}. Unfortunately, due to limited light from the sample compared to the camera frame rate requirements it is not suitable for ophthalmic imaging. The compromise that must be made between the extent of the field of view and the temporal resolution prevents LDF from imaging retinal blood flow with a temporal resolution sufficient to reveal the flow pulsatility. Finally, for the issue of imaging the direction of blood flow, LSCI and LDF share the limitation that for self-interfering light, positive and negative frequency shifts produce the same beat frequency, and therefore the same interferometric signal. As the Fourier transform of a real signal is symmetrical, LSCI and LDF cannot reveal the direction of blood flow.

Laser Doppler holography (LDH) is another optical Doppler imaging technique that can be described as the interferometric version of full-field LDF~\cite{SimonuttiPaquesSahel2010, Pellizzari2016}. The most notable conceptual difference between LDH and LSCI/LDF and also Doppler ultrasound is that a reference arm is introduced to measure the frequency shift by optical mixing. LDH thus benefits from an interferometric coherent gain, which ultimately enables the use of ultrafast CMOS cameras at frame rates sufficiently high to sample Doppler frequency shifts up to a few tens of kHz over a full-field of view. Blood flow in human retinal and choroidal vessels of a few tens of microns diameter can be imaged this way with a temporal resolution of a few ms~\cite{Puyo2018, Puyo2019, Puyo2019b}. Another valuable advantage of the holographic detection is to enable a digital compensation for optical aberrations~\cite{ColombMontfort2006, Kumar2013, Hillmann2016, Atlan2020}. Finally, what is explored in this manuscript is that the numerical holographic propagation also enables access to the phase of the optical field, and therefore recovering the sign of Doppler frequency shifts. We demonstrate what we believe to be the first full-field optical color Doppler imaging. We show that in out-of-plane vessels, the strong contribution of backscattering to the Doppler signal enables the local axial direction of blood flow to be revealed with high temporal resolution. We then show that this feature can be used to evidence arterial retrograde flow in pathological cases. This is to our knowledge the first time that human retinal blood flow reversal is imaged non-invasively and estimated quantitatively.

\section*{Setup and data processing}

\textbf{Optical setup} 
We used the Mach-Zehnder LDH setup presented previously~\cite{Puyo2018}. As a light source, we either used a $\SI{50}{\milli\watt}$ single-frequency laser diode operating at $\SI{785}{\nano\meter}$ wavelength (Thorlabs LP785-SAV50, VHG wavelength-stabilized), or a $\SI{20}{\milli\watt}$ single frequency laser diode operating at $\SI{852}{\nano\meter}$ (Thorlabs FPV852S, VHG wavelength-stabilized). Experimental procedures adhered to the tenets of the Declaration of Helsinki, and the study was approved by an ethical committee (Comit\'e de Protection des Personnes; clinical trial NCT04129021). Written informed consent was obtained from all subjects. The power of the laser beam incident at the cornea was $\SI{1.5}{\milli\watt}$ of constant exposure, and the eye fundus was most of the time imaged over a field of view of approximately 3 to 6 mm. The light backscattered by the retina interferes with an on-axis reference beam to form holograms which are digitally recorded by an ultrafast CMOS camera (Ametek - Phantom V2511, quantum efficiency 40\%, 12-bit pixel depth, pixel size $\SI{28}{\micro\meter}$) sampling at $f_{S} = 67 \; \rm kHz$ in a $512 \times 512$ format.

\textbf{Data processing} 
The data processing applied to the interferograms aims at revealing blood flow with a short-time Fourier transform analysis, with a typical sliding window size of 1024 holograms:
\begin{enumerate}
  \item The diffracted speckle pattern captured by interferograms is numerically propagated to the retinal plane by angular spectrum propagation. A phase kernel is applied in the spatial Fourier of each hologram~\cite{Goodman2005, Puyo2018}, and at this point the reconstructed holograms have become complex-valued.
  \item The short-time window is then filtered by singular value decomposition (SVD) to reject the Doppler signal from eye motion as detailed elsewhere~\cite{Puyo2020}. The SVD takes advantage of the strong spatial coherence over the field of view of Doppler contributions from global tissue motion. Local Doppler contributions from blood flow on the other hand, generate Doppler frequency shifts that are all uncorrelated throughout the field of view. Therefore, the projection of the holograms space-time matrix onto the eigenvectors associated with the eigenvalues of highest energy can be rejected to improve the access to blood flow contributions of slower velocities.
  \item The Doppler power spectrum density is computed from the squared magnitude of the temporal Fourier transform of the SVD-filtered hologram short-time window~\cite{Puyo2018}. It is denoted $S(x,y,f)$, where (x,y) are the pixel indices, and $f$ is the temporal frequency.
\end{enumerate}
The reconstructed holograms are complex valued, which implies that the positive and negative parts of the temporal Fourier transform are not necessarily symmetrical. The asymmetry stems from the fact that the frequency of a wave reflected by a moving object is positively or negatively Doppler shifted depending on whether the object is moving towards or away from the incoming wave. In order to characterize the spectral distribution of energy, different linear combinations of the power spectrum density are computed:
\begin{equation}  \label{eq:eq_Moments}
M_{0^{\pm}} = \int_{f_{c}}^{f_{N}} S(f) \pm  S(-f) \; \diff f, \quad
M_{1^{\pm}} = \int_{f_{c}}^{f_{N}} \left[ S(f) \pm  S(-f) \right] \; f  \; \diff f
\end{equation}
where $f_{c}$ is the lower frequency threshold generally set to 1 kHz, $f_{N} = f_{S}/2$ is the camera Nyquist frequency. $M_{0^{+}}$  and $M_{1^{+}}$ are said to be symmetrized in the sense that the two halves of the spectrum are folded in order to obtain a signal that is insensitive to the axial direction of blood flow. The zeroth moment $M_{0^{+}}$ is the area under the power spectrum curve, referred to as power Doppler. It is expected to represent the number of scatterers traveling at any speed above the equivalent velocity threshold, and therefore the quantity of moving red blood cells, i.e. the local blood volume~\cite{BonnerNossal1981, Bonner1990}. The first moment $M_{1^{+}}$ yields a quantity that takes into account both the number of photons above the frequency threshold and their frequency shift and is expected to represent blood flow.

\textbf{Color composite images}
Color composite images are obtained by merging two grayscale images $I_1$ and $I_2$ into a single colored image whose red/blue/green channels are computed in each pixel in the following way:
\begin{align*} 
    \rm{Red}   &=  c_1(1)\; I_1 + c_2(1) \; I_2\\
    \rm{Green} &=  c_1(2)\; I_1 + c_2(2) \; I_2\\
    \rm{Blue}  &=  c_1(3)\; I_1 + c_2(3) \; I_2
\end{align*}
where $c_1$ and $c_2$ are three element vectors that define two complementary colors. Composite images of slow/fast flow colored in cyan/red are obtained by fusing power Doppler images of low and high frequency ranges, generally 1-6 and 6-35 kHz. The contrast of the two power Doppler images is increased prior to the fusion. The resulting image is able to display an extended dynamic range of flow, and can help identify arteries in red and veins in cyan in the choroid due to their different velocities~\cite{Puyo2019}. The so-called systolodiastolic color composite images are similarly based on the fusion in orange/blue of power Doppler images that are temporally averaged over systole and diastole (with a manual selection of the time points). These images are useful to distinguish retinal arteries and veins thanks to their different pulsatility~\cite{Puyo2019b}. The areas where blood flow is greater during systole appear in orange, and those where blood flow is greater during diastole appear in blue.


\textbf{Spectrograms} 
In LDH, the baseline signal must be subtracted from the raw signal so that the measured local signal accounts for local blood flow.~\cite{Puyo2019b}. When calculating $M_{0^{+}}$ or $M_{1^{+}}$, the baseline signal to be subtracted is straightforwardly obtained by averaging the corresponding function over the whole field of view. When representing spectrograms or when estimating the mean frequency shift, the baseline subtraction must be carried out on the spectrum itself. This baseline subtraction calls for a prior correction of the inhomogeneities of intensity over the field of view (vignetting). Of the several ways to achieve this goal we chose to do as follows: starting from the power spectrum density matrix $S(x,y,f)$ previously filtered by SVD, an image is constituted by calculating the spectral average of $S(x,y,f)$, and blurred with a spatial Gaussian filter. The 3D power spectrum matrix $S(x,y,f)$ is divided by this image, and the total energy of the resulting matrix is re-normalized to ensure that the sum of the pixels energy is the same before and after the division by the blurred image. The power spectrum density is spatially averaged over the specific region of interest, which usually covers the whole lumen of a vessel to have a higher signal-to-noise ratio. Finally, the baseline spectrum, that is spatially averaged over the entire field of view, is subtracted from the measured spectrogram trace. The so-called symmetrized spectrograms are computed by a summation of the positive and negative parts of the spectrum. To enable the estimation of the mean frequency shift, the symmetrized spectrogram is set to 0 over the frequencies where it is negative due to the subtraction of the baseline spectrum. The mean frequency shift is estimated from the symmetrized spectrogram, and it is plotted in red and superimposed on the symmetrized spectrogram.

The mean frequency shift and the frequencies graduating spectrograms are converted into the equivalent velocities given by the Doppler formula $v_{\rm eq} = \lambda \Delta f  / {2 n \cos \alpha}$, where $\Delta f$ is the frequency shift, $\lambda$ is the wavelength, $n = 1.35$ is the refractive index, and $\alpha$ is the angle between the vessel and the optical axis. The value of the tuning parameter $\alpha$ is set such that $\cos \alpha = 0.5$ to take into account the variety of possible vessel angles whose cross-section is visible in the image, as well as the diversity of incident and scattered optical wavevectors contributing to the Doppler signal at one image point. With a sampling frequency of 70 kHz and wavelength of $\SI{785}{\nano\meter}$, the equivalent axial velocity corresponding to the Nyquist frequency 35 kHz (maximum that can be measured without aliasing) is $\SI[inter-unit-product = \ensuremath{{}\cdot{}}]{20}{\milli\meter\per\second}$. For a wavelength of $\SI{785}{\nano\meter}$, the 1 kHz threshold used in the computation of $M_{0^{\pm}}$ and $M_{1^{\pm}}$ correspond to an equivalent minimum detectable velocity of $\SI[inter-unit-product = \ensuremath{{}\cdot{}}]{0.6}{\milli\meter\per\second}$.


\section*{Directional contrast} \label{section_DirectionalContrast}

\begin{figure}[t!]
\centering
\includegraphics[width = 1\linewidth]{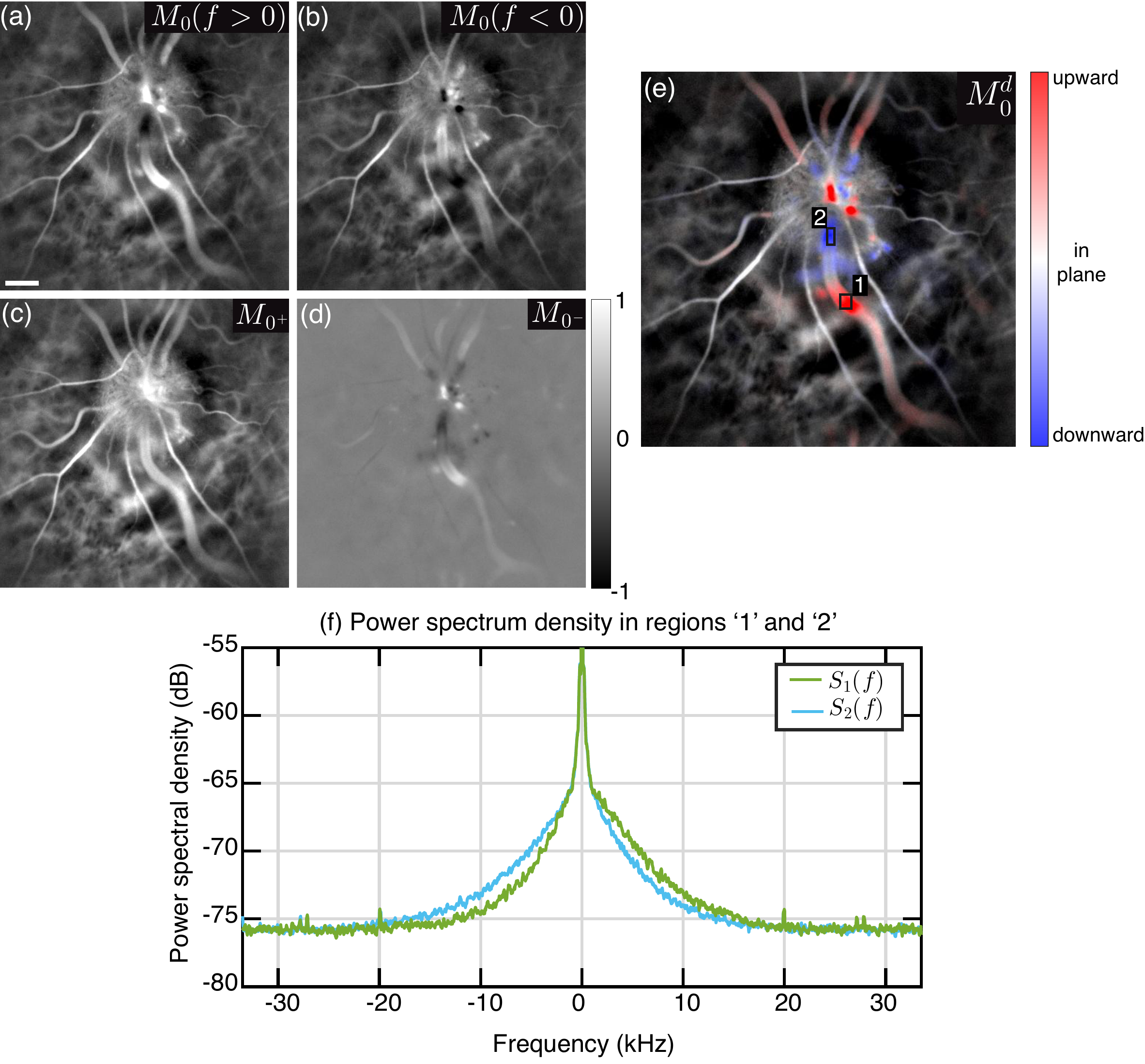}
\caption{Directional blood flow contrast. (a-b) Power Doppler images from the positive and negative parts of the Doppler spectrum. (c-d) Sum and difference of those two images. (e) Directional power Doppler image on which upward and downward flow give red and blue contrasts, respectively. (f) Doppler power spectrum density in two regions of a blood vessel where the local flow has upward (green trace) and downward components (blue trace), respectively.}
\label{fig_1_Spectrum}
\end{figure}

In a case of direct backscattering, if the scatterer is moving towards or away from the light beam, the induced Doppler frequency shift is positive or negative, respectively. The directional contrast we present here exploits the resulting Doppler spectrum asymmetry of light directly backscattered by flowing blood to reveal its axial direction. Power Doppler images of the optic disc region of a human retina for the positive and negative parts of the Doppler spectrum, denoted $M_0(f > 0)$ and $M_0(f < 0)$, are shown in Fig.~\ref{fig_1_Spectrum}(a) and (b). These two images visibly differ in some areas depending on the local axial direction of blood flow. In the areas where blood flow has a significant upward component, the spectral energy is greater on the positive side, and conversely when blood flow has a significant downward projection, there is more energy in the negative part of the spectrum. The sum and difference of $M_0(f > 0)$ and $M_0(f < 0)$ are computed to obtain $M_{0^{+}}$ and $M_{0^{-}}$, shown in Fig.~\ref{fig_1_Spectrum}(c) and (d), respectively. The symmetrized power Doppler image $M_{0^{+}}$ reveals all blood flow structures, whereas the differential spectral image $M_{0^{-}}$ shows signal only in some areas. Its value is negligible where blood vessels are perpendicular to the optical axis, and it is positive and negative where blood flow has a significant upward or downward projection, respectively. The purpose of the directional contrast is to combine the information of both images in a single one. To that end, a hue/saturation/value (HSV) image combination scheme is used based on $M_{n^{-}}$ and $M_{n^{+}}$ images, with $n$ indifferently equal to 0 or 1 (zeroth or first moment of the power spectrum density). The three matrices used for the HSV composition at pixel indices $(x,y)$ are the following:
%
\begin{align*}
    \rm{Hue}&= 
    \begin{cases}
      \rm {red}, & \text{if}\ M_{n^{-}} > 0 \\
      \rm {blue}, & \text{otherwise} 
    \end{cases} \\
    \rm{Saturation} &= \lvert {M_{n^{-}}} \rvert \\
    \rm{Value} &= M_{n^{+}}
\end{align*}
%
The hue matrix is obtained by binarization of $M_{n^{-}}$, it results in a red color where $M_0(f > 0) > M_0(f < 0)$, and a blue color otherwise. The saturation matrix is obtained from the absolute value of $M_{n^{-}}$, and the value matrix is simply the image $M_{n^{+}}$. The images resulting from this HSV combination for the zeroth and first moment are called directional blood volume and blood flow images, denoted $M^{d}_0$ and $M^{d}_1$, respectively. The purpose is to color the intensity of $M_{n^{+}}$ images to have flow towards the camera depicted in red, and flow away from the camera depicted in blue. The color saturation depends on the spectral asymmetry: the image remains grayish (low saturation) where the Doppler spectrum is approximately symmetrical and becomes more vividly colored (high saturation) where the asymmetry is greater. An example of a directional blood volume image is given in Fig.~\ref{fig_1_Spectrum}(e), where the red/blue contrast based on the spectral asymmetry is added to the usual power Doppler image. Two regions of interest in areas with red and blue contrasts along a single vein are considered. The power spectrum density in each region is plotted in Fig.~\ref{fig_1_Spectrum}(f) for a given short-time window. The power spectrum density in region '1' is shifted towards positive frequencies, and it is the other way around in region '2', where the energy is shifted towards negative frequencies. This is due to the difference of flow direction with respect to the optical axis in both regions, which gives a hint about the local topography of the tissue supporting the observed vessel. Blood flow has an upward component in region '1' because the vessel goes over the optic nerve head (ONH) ridge, then blood flow has a downward flow component in region '2' as the vein plunges into the emergence of the central retinal vein.

\begin{figure}[t!]
\centering
\includegraphics[width = 1\linewidth]{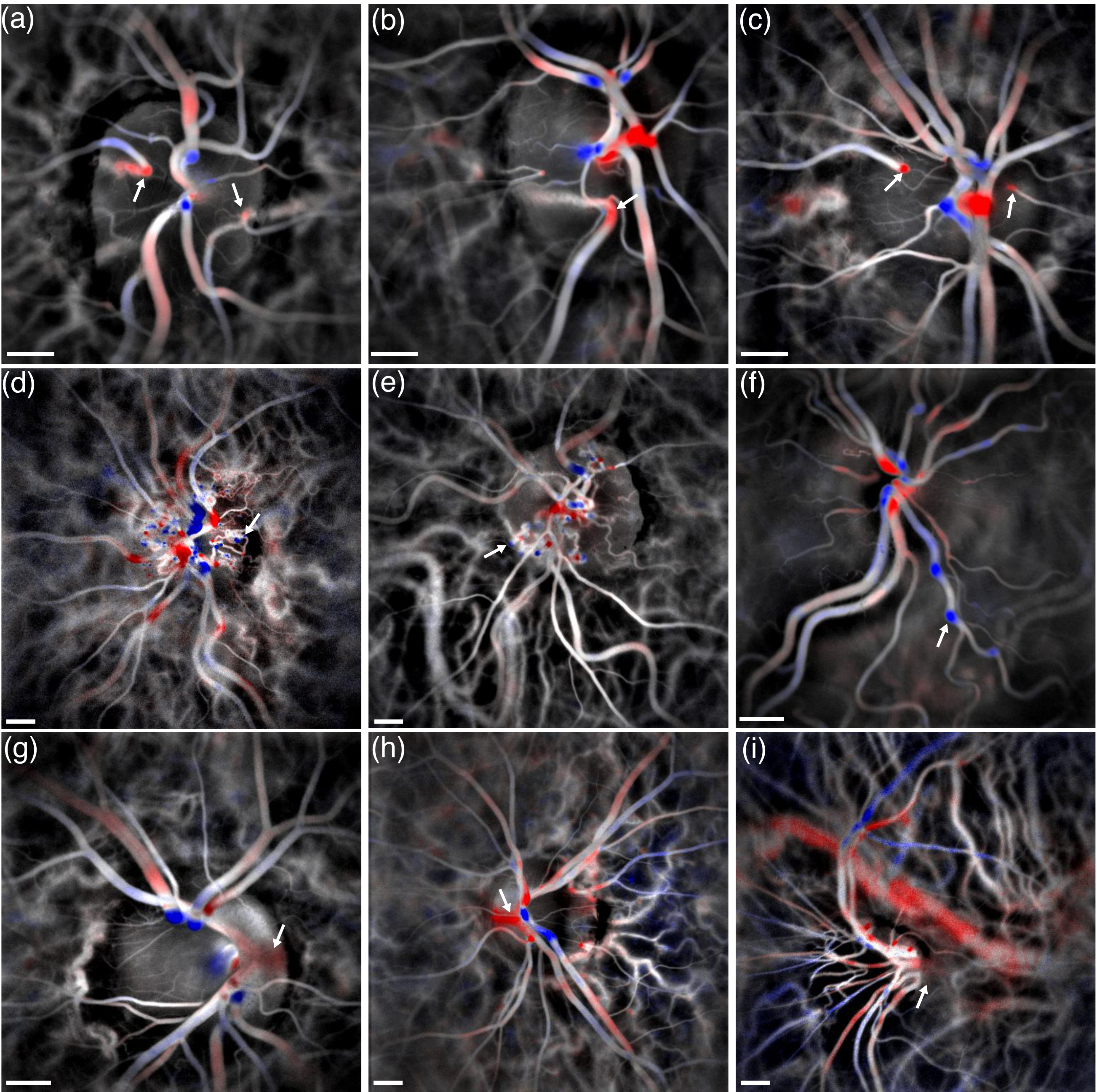}
\caption{Examples of vascular structures revealed with the directional contrast. (a-c) Flow incoming from cilio-retinal arteries. (d-f) Axial tortuosity in collateral or large retinal vessels. (g-i) Abnormal inclination of the papilla and central retinal vessels (papillary dysversion). The estimated size of the scale bar is $\SI{500}{\micro\meter}$.
}
\label{fig_2_DirectionalExamples}
\end{figure}

Examples of directional LDH images in healthy and pathological eyes are shown in Fig.~\ref{fig_2_DirectionalExamples}. The first three examples in Fig.~\ref{fig_2_DirectionalExamples}(a-c) show incoming flow from cilio-retinal arteries that can be readily identified from the red contrast. Cilio-retinal arteries are supplied from branches of posterior ciliary arteries located deeper in the ONH and typically present an upright section that makes them visible with the directional contrast. Cilio-retinal arteries are especially visible as they are well separated from the rest of retinal vasculature of the optic disc. As shown in Fig.~\ref{fig_2_DirectionalExamples}(b), even the lumen of surprisingly small cilio-retinal arteries can be made visible with the directional contrast.
Secondly, Fig.~\ref{fig_2_DirectionalExamples}(d-f) shows images where the axial tortuosity in vessels can be observed. Collateral vessels that have formed in the optic disc to shunt retinal vein occlusions are seen in Fig.~\ref{fig_2_DirectionalExamples}(d) and (e). The venous flow circulates in these entangled vessels and the sections where the flow has a strong axial component are revealed. The directional image in Fig.~\ref{fig_2_DirectionalExamples}(f) shows an example of tortuosity in large retinal vessels, where once again the sections of vessels where flow has a significant axial component are directly identified.
Finally, the examples provided in Fig.~\ref{fig_2_DirectionalExamples}(g-i) are cases of abnormal orientation of the papilla and central retinal vessels (papillary dysversion). In Fig.~\ref{fig_2_DirectionalExamples}(g) and (h), the spreading of the red contrast highlighted by arrows indicates a mild inclination of the emergence of the central retinal artery. The blurriness is presumably due to the overlying nerve fiber layer.
In Fig.~\ref{fig_2_DirectionalExamples}(i), the angle between the optical axis and the emergent central retinal artery and vein is so large that there is not even red or blue contrast to identify them.

Overall, although it does not offer a quantitative value for the flow axial orientation, the directional contrast is a useful tool to give a better understanding of how the flow circulates axially. It is an additional feature that compensates for the lack of 3D information of LDH images. It is particularly helpful to determine the entrance and exit points of blood flow into the retina and identify the emergence of the central retinal artery and vein, and of cilio-retinal arteries. This is however not systematically possible, as for example when the vessels are entangled or emerge from the optic disc with a horizontal inclination. All the images presented in Fig.~\ref{fig_2_DirectionalExamples} were obtained using the 1-33 kHz frequency range. As previously explored~\cite{Puyo2018}, smaller blood vessels are better revealed by using lower frequency ranges, but global eye movements also have a response range in the low frequencies. These contributions may sometimes not be fully suppressed by the SVD and are particularly unwelcome when using the directional contrast because the oscillating axial motion of the eye appears with wide-field red and blue flashes (data not shown).

\begin{figure}[t!]
\centering
\includegraphics[width = 1\linewidth]{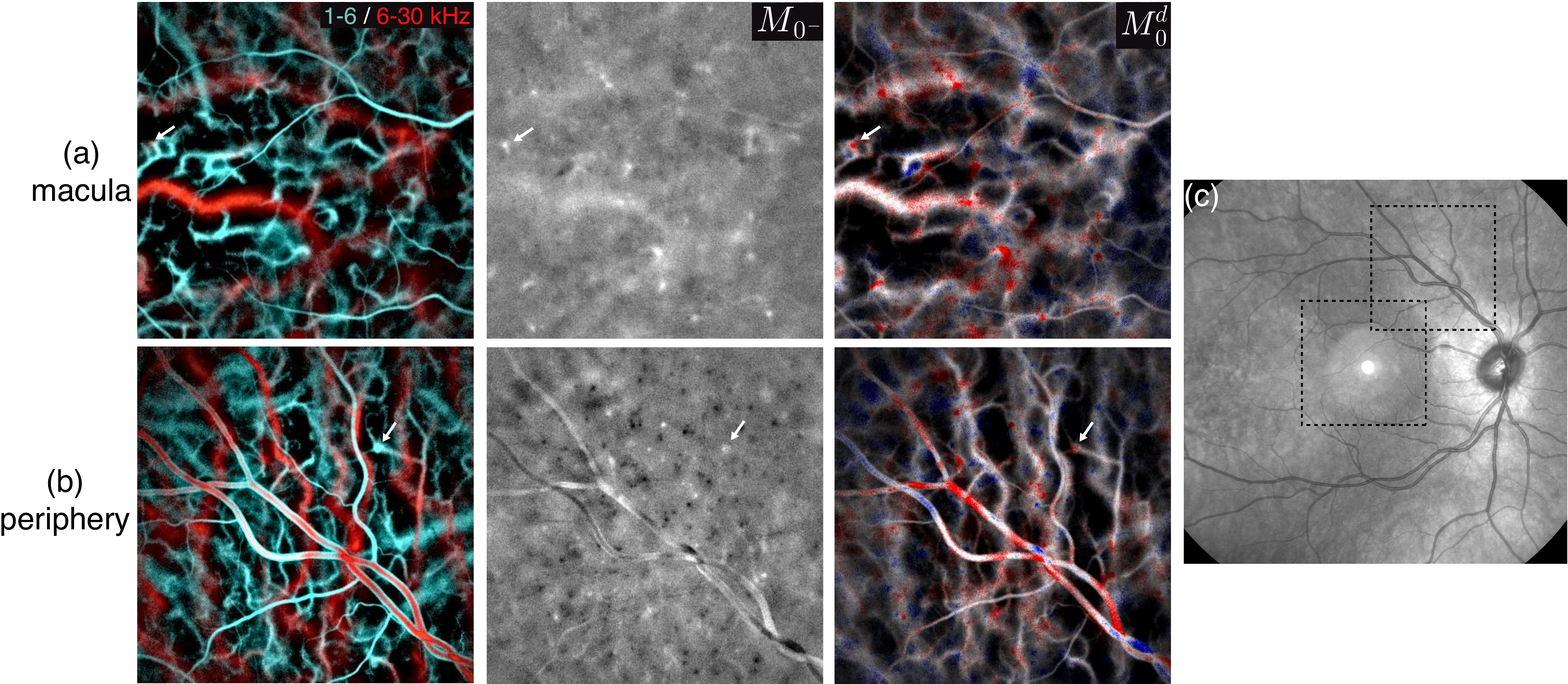}
\caption{
Directional contrast in choroidal structures. (a-b) In a macular and peripheral region, the slow/fast flow color composite images, the spectrum difference, and the directional power Doppler images are shown. The upright sections of the choroidal arterioles and venules (white and black dots) are better revealed in the peripheral region. (c) Scanning laser ophthalmoscope image showing the two regions.
}
\label{fig_3_ChoroidExamples}
\end{figure}

In Fig.~\ref{fig_2_DirectionalExamples}, we observe that the directional contrast is higher in retinal vessels than in choroidal vessels. For example, many choroidal arteries surrounding the optic disc are expected to be axially oriented since they are branches of posterior ciliary arteries, which are located posteriorly. However, there is little to no directional contrast visible in these vessels. The differential contrast between the two halves of the Doppler spectrum is reduced because light from the choroid suffers from multiple scattering, due to the epithelial layer, choroidal melanin, or photoreceptor mosaic. The directional contrast relies on backscattered light whose direction has not been randomized prior to being scattered by blood cells, thus the lower ratio of singly scattered versus multiply scattered light in the choroid hinders the directional contrast in choroidal vessels. To illustrate the directional contrast in the choroid, images of a macular and a peripheral region in the same eye are shown in Fig.~\ref{fig_3_ChoroidExamples}(a-b). The location of the two imaged regions is shown on a scanning laser ophthalmoscope image in Fig.~\ref{fig_3_ChoroidExamples}(c). For each measurement the slow/fast flow color composite image, the spectral difference image, and the corresponding directional power Doppler image are shown. The spectral difference image enables better visualization of the spectral asymmetry in choroidal vessels of small blood flow than the directional blood flow image. In the macular region, except for a few white dots, little directional contrast is seen. In particular, almost no directional signal is seen in the large posterior ciliary artery in the middle of the field of view. In the peripheral region, more black and white dots are visible. These dots most likely indicate the position of the arterioles and venules of the choroid that present an upright section as they connect the superficial choriocapillaris layer of the choroid to the deeper layer where the largest choroidal vessels are located. The improvement in directional contrast with the eccentricity is likely due to the lower photoreceptor and/or epithelial cell densities~\cite{Bhatia2016, Liu2017noninvasive}, or possibly from the thinner choroid~\cite{Hoseini2019}.

\section*{Spectrograms} \label{Spectrograms}

\begin{figure}[t!]
\centering
\includegraphics[width = 1\linewidth]{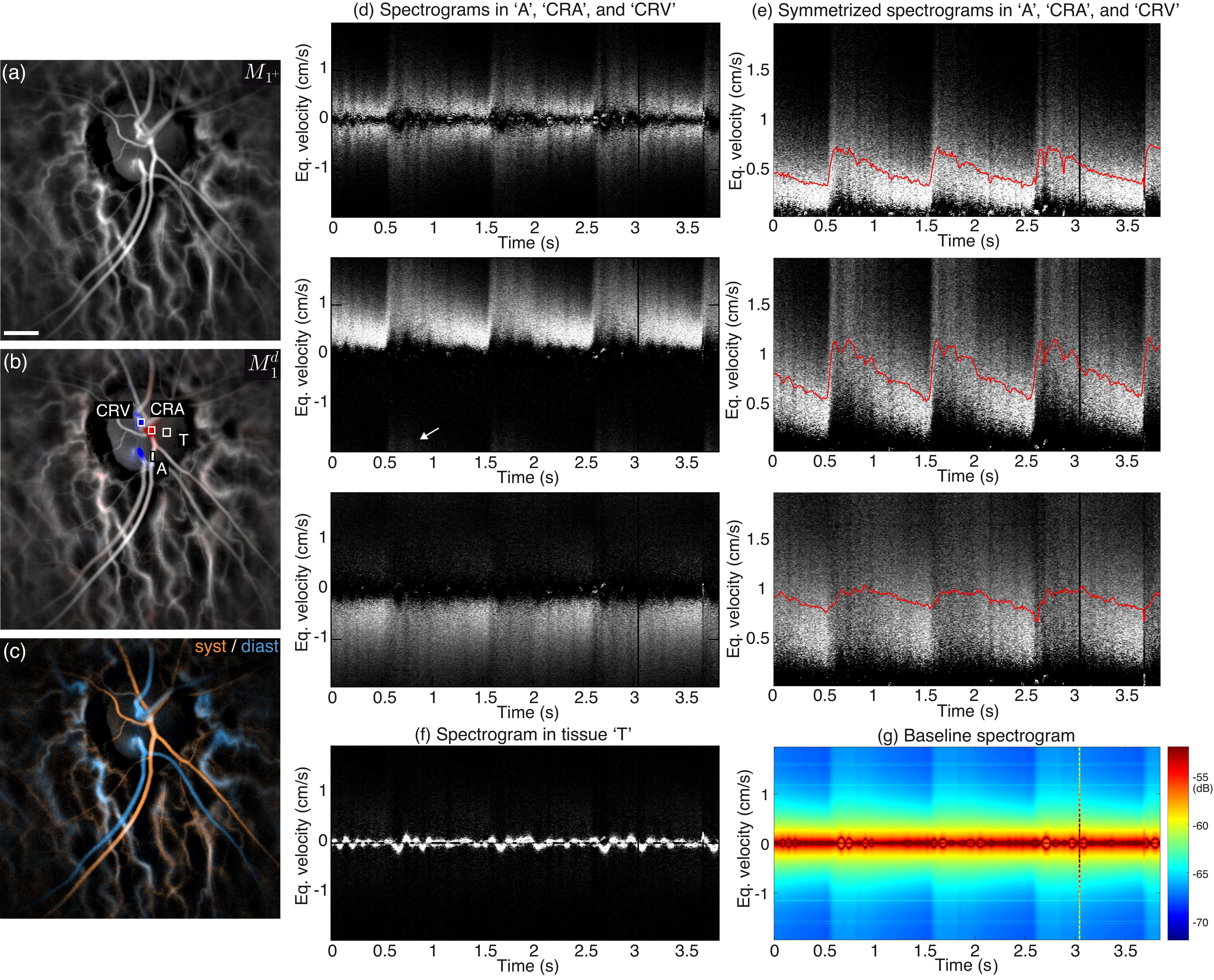}
\caption{Spectrographic flow measurements (healthy subject). (a-b) Blood flow and directional blood flow. (c) Color composite image of systolodiastolic blood flow changes. (d) Spectrograms in an in-plane artery ('A') where multiply scattered light contributes to the Doppler spectrum, and in the emergence of the central retinal artery ('CRA') and central retinal vein ('CRV') where backscattered light is the main contribution the Doppler spectrum. (e) Symmetrized spectrograms in the same regions. (f) The spectrogram in the region 'T' shows the ONH tissue motion. (g) Baseline spectrogram (spatial average). \textcolor{blue}{\href{https://youtu.be/sTaTKAsoNAs}{Visualization 1}} shows the directional blood flow movie juxtaposed to the spectrogram in the 'CRA' region. The estimated size of the scale bar is $\SI{500}{\micro\meter}$.
}
\label{fig_3_CRA_CRV_EKL}
\end{figure}

In Fig.~\ref{fig_3_CRA_CRV_EKL}, we use spectrograms to investigate the Doppler spectrum asymmetry and its pulsatile changes in vessels with different axial orientations. In Fig.~\ref{fig_3_CRA_CRV_EKL}(a) and (b) are shown the blood flow and directional blood flow images of a healthy subject, i.e. $M_{1^{+}}$ and $M^{d}_1$, respectively. A systolodiastolic color composite image based on blood flow variations is shown in Fig.~\ref{fig_3_CRA_CRV_EKL}(c) and is able to differentiate arteries and veins in orange and blue. As visible in \textcolor{blue}{\href{https://youtu.be/sTaTKAsoNAs}{Visualization 1}}, the red/blue contrast in individual directional blood flow images is sufficient to differentiate arteries from veins. This makes the directional contrast a more powerful way of differentiating arteries from veins than systolodiastolic images; firstly because the intensity (i.e. the value component of the HSV composition) of directional blood flow images remains as quantitative as the usual blood flow images, which is not the case of systolodiastolic images due to the contrast adjustment; secondly because the directional contrast offers the possibility to identify arteries and veins directly on blood flow movies, which is much more convenient for clinicians than having to refer to a second image while watching the blood flow movie, and finally, unlike systolodiastolic images, directional images do not require a manual intervention.

Thanks to the directional blood flow image in Fig.~\ref{fig_3_CRA_CRV_EKL}(b), we identified an area 'A' in the optic disc positioned on an in-plane retinal artery (grayish directional contrast). Two other areas are positioned on the emergence of the central retinal artery ('CRA') and vein ('CRV'), where blood flow has a large upward and downward out-of-plane component (visible from the strong directional contrast), respectively. The spectrograms in these three regions of interest are shown in Fig.~\ref{fig_3_CRA_CRV_EKL}(d). It should be noted that the baseline spectrogram shown in Fig.~\ref{fig_3_CRA_CRV_EKL}(g) has been subtracted from all the local spectrograms represented in shades of gray. The first two spectrograms both show Doppler spectra in vessels of arterial type, but they present dramatically different energy distributions. In region 'A' where blood flows in-plane, the energy of the spectrogram is equally distributed between the positive and negative frequencies, which is expected as only multiply scattered light contributes to the Doppler signal. Indeed, as the Doppler angle for in-plane blood flow yields a negligible Doppler frequency shift on directly backscattered light, it is only thanks to multiple scattering that LDH can be sensitive to in-plane blood flow. In the 'CRA' region where blood flows upward, the energy of the spectrogram lies however almost exclusively in the positive frequencies, and in the 'CRV' region where blood flows downward, the energy is located in the negative part of the spectrum. If there was multiply scattered light, it would contribute to both the negative and positive parts of the Doppler spectrum. Therefore, one can conclude from the asymmetrical spectrum in the 'CRA' and 'CRV' regions that single backscattering is the dominant contribution to the Doppler spectrum in axially oriented vessels. This weak contribution of multiply scattered light to the Doppler spectrum in axially oriented retinal vessels explains why the directional contrast remains strong in individual blood flow images as seen in \textcolor{blue}{\href{https://youtu.be/sTaTKAsoNAs}{Visualization 1}}.

For in-plane blood flow, because the energy of spectrograms is symmetrically distributed between positive and negative frequencies, the estimation of the average frequency would be 0 for any flow velocity. Therefore, the local average frequency shift is measured instead with symmetrized spectrograms. The corresponding symmetrized spectrograms for the three same regions of interest and averaged frequency shifts (red plots) are shown in Fig.~\ref{fig_3_CRA_CRV_EKL}(e). This time a similar arterial flow profile is observed in the 'A' and 'CRA' regions, and a venous flow profile is seen in the 'CRV' region. The exact orientation of the vessels with respect to the optical axis is unknown, so we cannot correct for the influence of the angle between the scattering wave vectors and blood flow direction and the present spectrograms are only graduated in equivalent velocity. The equivalent velocities measured on these spectrograms are however in the range of what would be expected from instruments able to make absolute velocity measurements. With Doppler-OCT, the mean velocity measured in retinal vessels in the vicinity of the optic disc is around a few cm/s~\cite{Baumann2011, Werkmeister2012, Haindl2016, Wartak2016, Desissaire2020}. A final comment is that one can see in the 'CRA' spectrogram that a part of the Doppler spectrum is aliased during systole due to temporal undersampling (arrow). Aliasing can be problematic by hindering the measurement of the average frequency shift. In this specific case, it seems to have only a minor effect, as the blood velocity profile in the 'A' and 'CRA' are quite similar except at the systolic peak.

Finally, the spectrogram in Fig.~\ref{fig_3_CRA_CRV_EKL}(f) shows the Doppler response of the ONH tissue, measured in the area 'T'. No high frequency blood flow signal can be seen, but there is a very low frequency oscillating pattern caused by the axial motion of the eye~\cite{Singh2010, Puyo2021a}. The baseline spectrogram shown in Fig.~\ref{fig_3_CRA_CRV_EKL}(g) is the spectrogram spatially averaged over the whole field of view. This baseline integrates all contributions: retinal and choroidal blood flow, as well as tissue motion. It is interesting to observe that in contrast to the spectrogram in tissue 'T', the Doppler traces at very low frequency (bright red) are symmetrical. This is because when averaging over the entire field of view, the signal of both holographic twin images is integrated, and one carries the complex conjugate signal of the other. Therefore, the holographic twin image contributes to the signal despite being completely out-of-focus, and it gives the symmetrical signal of tissue motion.

\begin{figure}[t!]
\centering
\includegraphics[width = 1\linewidth]{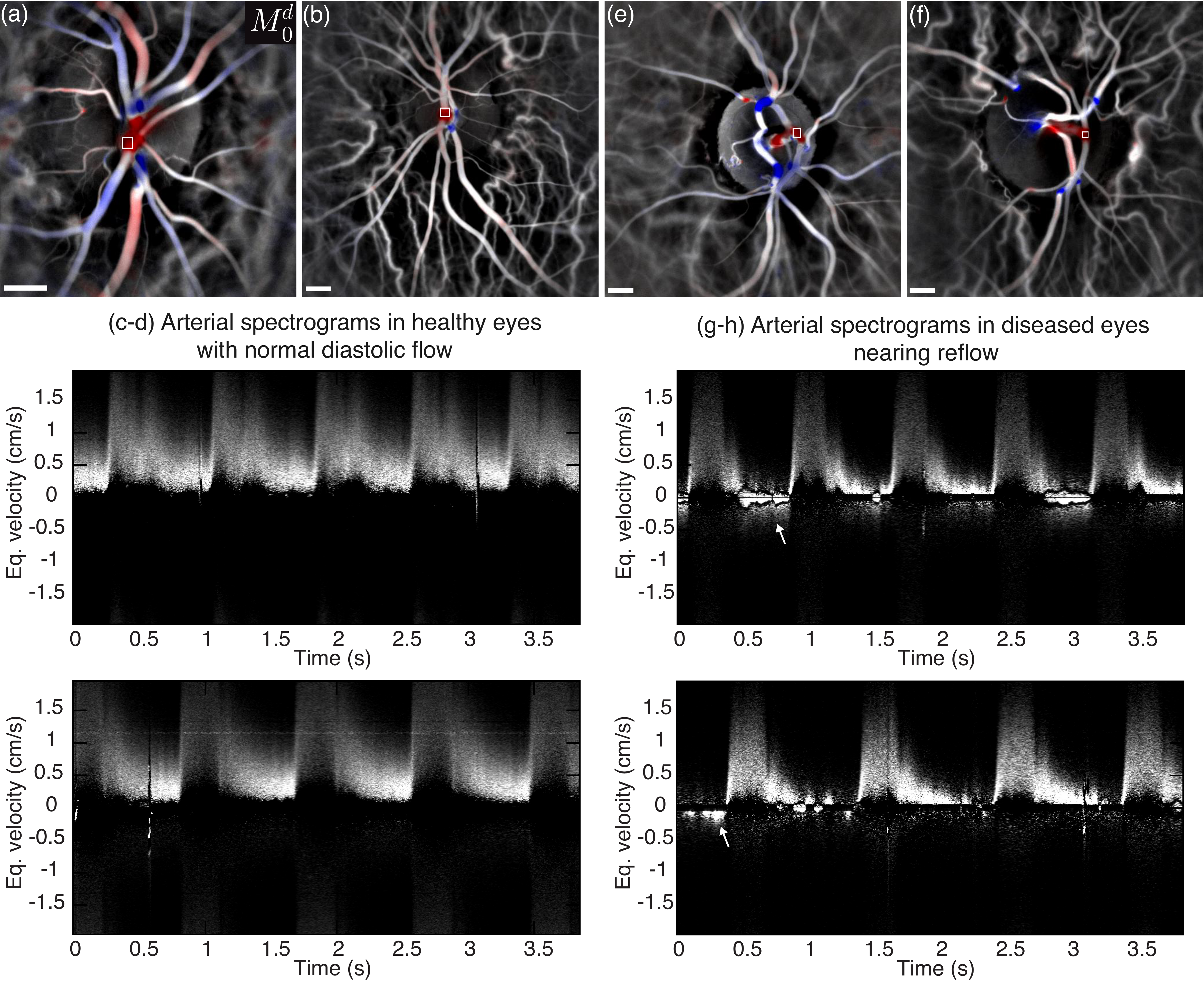}
\caption{
Example of spectrograms in healthy and diseased eyes. (a-b) In the two healthy eyes the arterial flow is anterograde throughout the whole cardiac cycle. In these cases of (c) central retinal vein occlusion and (d) high tension glaucoma, the diastolic arterial flow is below the limit where LDH can reliably monitor blood flow variations. The estimated size of the scale bar is $\SI{500}{\micro\meter}$. \textcolor{blue}{\href{https://youtu.be/qXn64v5sj8E}{Visualization 2}}.
}
\label{fig_new_spectrograms}
\end{figure}

Figure \ref{fig_new_spectrograms} shows other examples of spectrograms measured in the emergence of the central retinal artery of healthy and diseased eyes. For each eye the directional power Doppler image and the spectrogram measured in an out-of-plane artery are shown. \textcolor{blue}{\href{https://youtu.be/qXn64v5sj8E}{Visualization 2}} shows the directional blood flow movies with the exact region of interest for all eyes, juxtaposed with the corresponding spectrograms. In the two healthy eyes featured in Fig.~\ref{fig_new_spectrograms}(a-b), the arterial blood flows throughout the whole cardiac cycle, i.e. the ocular perfusion pressure remains sufficiently high during diastole to overcome the vascular resistance. This is also visible on the corresponding spectrograms in Fig.~\ref{fig_new_spectrograms}(c-d). The other two eyes shown in Fig.~\ref{fig_new_spectrograms}(e) and (f) were affected by central retinal vein occlusion, and high-tension glaucoma, respectively, which both amount to an increase in vascular resistance. The spectrogram reveals that the arterial flow becomes so slow during diastole that it can fall below the detection threshold of LDH or can be difficult to distinguish from the contribution of eye motion. The energy of the corresponding spectrograms in Fig.~\ref{fig_new_spectrograms}(g-h) is even carried by negative frequencies at some moments(arrows), which could be the sign of a retrograde flow. However, because the negative signal is not recurrent over consecutive cardiac cycles, it is not presently possible to say if it is caused by an irregular and small flow reversal, or an artifact caused by the axial eye movement.

\section*{Blood flow reversal} \label{section_Reversal}

\begin{figure}[t!]
\centering
\includegraphics[width = 1\linewidth]{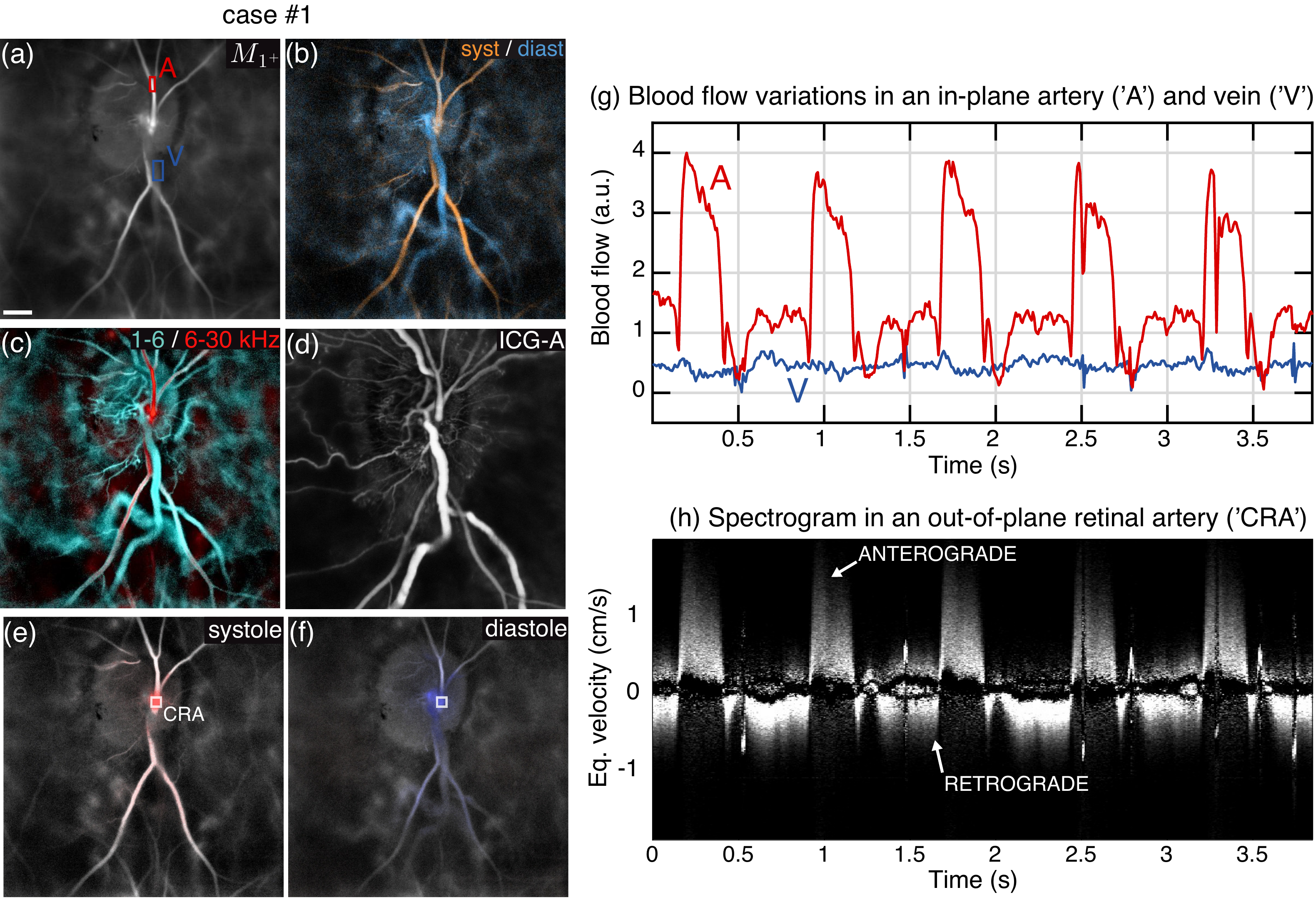}
\caption{Blood flow reversal in the central retinal artery in a case of high tension glaucoma/CRVO. (a) Averaged blood flow image. (b) Systolodiastolic image. (c) Color composite Doppler image of low/high flow in red/cyan. (d) ICG-Angiogram. (e-f) Directional blood flow during systole and diastole. (g) Blood flow measured in an in-plane artery 'A' and vein 'V'. (h) Spectrogram measured in 'CRA'. \textcolor{blue}{\href{https://youtu.be/94tQnibQG20}{Visualization 3}} shows the directional blood flow movie juxtaposed to the arterial spectrogram. \textcolor{blue}{\href{https://youtu.be/fGvFCZ7nTf4}{Visualization 4}} shows the dynamic ICG-angiography. The estimated size of the scale bar is $\SI{500}{\micro\meter}$.
}
\label{fig_6_Reflux}
\end{figure}

\begin{figure}[t!]
\centering
\includegraphics[width = 1\linewidth]{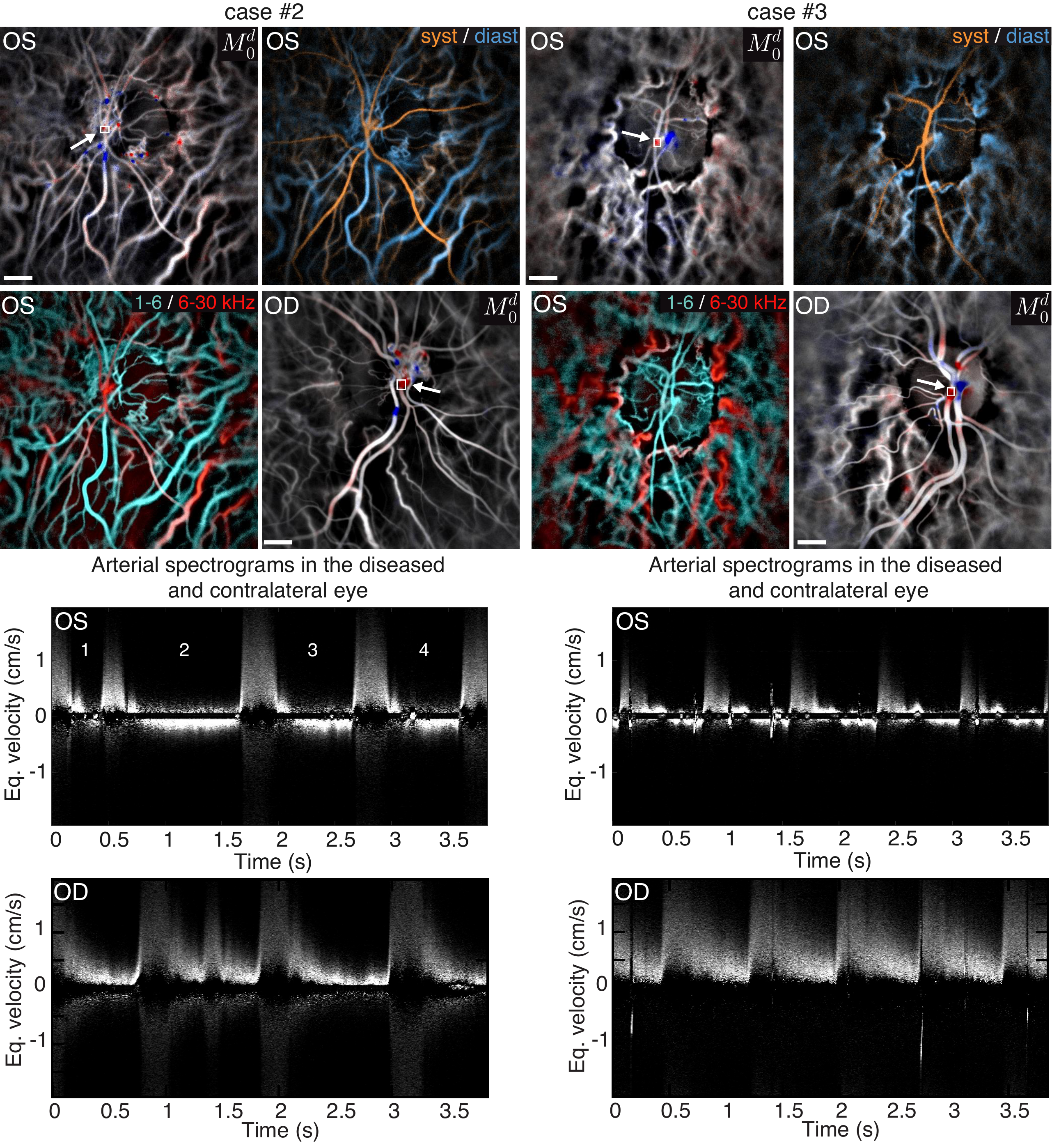}
\caption{Two other cases of arterial blood flow reversal. On the left is an eye affected by CRVO and glaucoma. On the right is a case of retinal vasculitis. For each case the color composite systolodiastolic and slow/fast flow images, and the directional power Doppler image are shown for the affected left eye (OS). The spectrogram measured in the emergence of the central retinal artery is shown for each case. The directional power Doppler and the arterial spectrograms are also shown for the contralateral right eye (OD). For both cases, the pathological increased vascular resistance causes a reflow. \textcolor{blue}{\href{https://youtu.be/YArAX_1o4oQ}{Visualization 5}} and \textcolor{blue}{\href{https://youtu.be/DXo5VWo4y2M}{Visualization 6}} show the directional blood flow movies juxtaposed to the spectrograms for the two eyes of both subjects. The estimated size of the scale bar is $\SI{500}{\micro\meter}$.
}
\label{fig_8_RefluxExamples}
\end{figure}

Finally, we present cases of distinct blood flow reversals in this section. The first example shown in Fig.~\ref{fig_6_Reflux} was found in an eye affected by high tension glaucoma (54 mmHg of IOP) and central retinal vein occlusion (CRVO). The averaged blood flow image is shown in Fig.~\ref{fig_6_Reflux}(a), and blood flow variations in an in-plane artery (A) and vein (V) are plotted in Fig.~\ref{fig_6_Reflux}(g). The composite systolodiastolic image revealing retinal arteries/veins in orange/blue is shown in Fig.~\ref{fig_6_Reflux}(b). The slow/fast color composite image obtained by fusing the 1-6 kHz and 6-33 kHz power Doppler images in cyan/red is shown in Fig.~\ref{fig_6_Reflux}(c). The vein is abnormally dilated, as typical in cases of retinal vein occlusions, so the venous blood flow appears significantly slower than the arterial blood flow. Therefore, on this composite image the retinal vein is seen in cyan and the artery in red. In Fig.~\ref{fig_6_Reflux}(e-f) the directional blood flow images averaged during systole and diastole are shown. It can be seen that the directional contrast in the emergence of the central retinal artery changes from red to blue over time, indicating that the direction of blood flow changes throughout cardiac cycles. The spectrogram in Fig.~\ref{fig_6_Reflux}(h) measured in the emergence of the central retinal artery reveals precisely how the flow changes direction throughout cardiac cycles. Unlike in previous examples, this time the energy of the arterial spectrogram does not remain over positive frequencies during the whole cardiac cycle. During systole, the energy is carried by positive frequencies as the arterial blood flow is anterograde. However, during diastole the energy is shifted to negative frequencies, which means the flow becomes retrograde. \textcolor{blue}{\href{https://youtu.be/94tQnibQG20}{Visualization 3}} shows the directional blood flow movies juxtaposed with the arterial spectrogram.

The same eye was also imaged with indocyanine-green angiography (ICG-A) with a Spectralis (Heidelberg Engineering). The ICG angiogram at the venous filling stage over the same field of view is presented in Fig.~\ref{fig_6_Reflux}(d). By itself, this image does not give much information about the flow, however the movie of the early circulation of the dye in the retina shown in \textcolor{blue}{\href{https://youtu.be/fGvFCZ7nTf4}{Visualization 4}} also seems to be able to reveal the arterial flow reversal, although not as unequivocally as LDH. It can be seen in the first few seconds of this movie that when the dye first arrives in retinal arteries, there are unusual pulsatile variations of the ICG fluorescence signal. In light of the demonstration of the pulsatile backflow observed with LDH, we know that these oscillations are caused by diastolic backflows: the systolic wave carrying the front wave of ICG goes forward and backward in the arterial vascular tree at the rhythm of the cardiac cycle.

It is interesting to study the non-directional blood flow variations measured in in-plane vessels shown in Fig.~\ref{fig_6_Reflux}(g), which stand in good agreement with the variations observed on the arterial spectrogram. The arterial flow profile is very unusual in comparison with controls, as after the systolic peak, blood flow falls to zero, makes a small bounce, and then a second more significant bounce until the diastole end. This early cancellation of blood flow followed by an increasing diastolic flow is exceptional, and thanks to the directional measurements, we know that these bounces are retrograde flow. A second observation is that the venous blood flow is also significantly lower because of the venous occlusion. However, it should also be considered that the total blood flow actually entering the retina is quite limited due to the flow reversal, therefore the flow exiting the retina must also be low. From the spectrogram, it is possible to compute the relative differential flow (RDF) entering from the central retinal artery such as (flow forward - flow backward)/(flow forward + flow backward), i.e. :
\begin{equation}  \label{eq:eq_RelativeDifferentialFlow}
RDF =  \dfrac{{\displaystyle \int_{-f_{N}}^{f_{N}} S(f) \, f \; \diff f}}{\displaystyle \int_{-f_{N}}^{f_{N}} S(f) \, \lvert f \rvert\ \;  \diff f}
\end{equation}
For this subject, we calculate a relative differential flow of 0.4.

Two other examples of blood flow reversal are shown in Fig.~\ref{fig_8_RefluxExamples}. As for the previous example, we show the images for the blood flow, the systolodiastolic changes in orange/blue, the slow/fast flow in cyan/red, the directional power Doppler, and the spectrogram measured in the emergence of the central retinal artery. In both cases, it was the left eye (OS) that was affected, but we also show the directional power Doppler and the arterial spectrogram in the contralateral right eye (OD).
On the left side of Fig.~\ref{fig_8_RefluxExamples}, images from an eye that was affected by both CRVO and glaucoma are shown. A diastolic blood flow reversal can be distinctly observed on the spectrogram, although it is not as dramatic as for the previous case. The subject also suffered from arrhythmia and had an extrasystole at the beginning of the measurement. It is interesting to note that the first two heartbeats are so close that there is no time for the reflow to begin (1). The heartbeat that follows is then longer than normal (2), but the maximum velocity of the reflow seems to remain similar to the velocity observed in the last two cardiac cycles (3 and 4). For this subject, we calculate a relative differential flow of 0.9. The contralateral eye also had had problems of retinal vein occlusions and shows signs of a high vascular resistance, but it did not present flow reversal. It can be observed on the spectrogram of the right eye that multiply scattered light due to the vessel not being exactly axially oriented generates a signal in negative frequencies. In \textcolor{blue}{\href{https://youtu.be/YArAX_1o4oQ}{Visualization 5}}, the directional blood flow movies and the arterial spectrogram are shown for each eye; the exact region of interest where the spectrogram is measured is indicated on the movie.
Finally on the right side of Fig.~\ref{fig_8_RefluxExamples} a case of retinal vasculitis is shown. For this subject, the retrograde flow is hard to discern on the directional blood flow movie in \textcolor{blue}{\href{https://youtu.be/DXo5VWo4y2M}{Visualization 6}} because the retinal blood flow in this subject was itself quite slow even during systole, especially compared to choroidal blood flow. The flow reversal is however very well visible on the spectrogram, and we calculated a relative differential flow of 0.75. In this subject, the contralateral right eye showed a normal arterial flow during diastole.

\section*{Discussion and conclusion}

In this manuscript, we have introduced an optical implementation of color Doppler imaging suited for LDH. We showed that this directional contrast allows us to make sense of the 3D geometry of vessels and identify the entrance and exit points of retinal blood flow in the ONH area and may also be useful to image choroidal arterioles and venules. It is interesting to note that the information about the axial direction of flow is retrieved from the complex optical field despite the presence of the defocused holographic twin image, which is yet expected to disrupt the phase measurements. We used LDH sensitivity to the axial direction of blood flow to evidence cases of retinal blood flow reversals in human subjects. We found retrograde flows in the retinal arteries of two eyes affected by both CRVO and glaucoma, and in a third eye affected by retinal vasculitis. The eye in which we observed the strongest arterial blood flow reversal in Fig.~\ref{fig_6_Reflux} also presented a high IOP (54 mmHg), and both the elevated IOP and the venous occlusion are expected to take part in the flow reversal by increasing the downstream vascular resistance and reducing the ocular perfusion pressure. These results are highly consistent with previous studies since retinal blood flow reversal had previously been reported in cases of human retinal vein occlusion~\cite{Paques2001, Paques2002, Paques2005}, and animal models of both elevated IOP and branch retinal vein occlusion~\cite{Dollery1968, Ffytche1974, PiJia2019, Peyman1998}. The occurrence and the effect of such arterial blood flow reversals remain to be investigated. Since they are caused by an abnormal downstream or upstream vascular resistance, blood flow reversals are a likely biomarker of poor retinal perfusion. The volume of oxygenated blood entering the retina is lowered because of the reversal, so it can be hypothesized that the retinal oxygenation and metabolism may be more hindered as the flow reversal becomes more significant. However, the difference between a case of backflow and a case where blood simply stops flowing should be elucidated. In the case of reversed flow in the ophthalmic artery, it has been found that it is a significant risk factor for ocular ischemic syndromes~\cite{Costa1997}.

The fact that epidemiological studies found that a lower diastolic blood pressure is dramatically associated with glaucoma occurrence could be related to the existence of recurrent diastolic retrograde retinal blood flows in those eyes~\cite{Tielsch1995}. Indeed, a lower diastolic blood pressure would make the retinal perfusion more vulnerable to diastolic blood flow reversal by further decreasing the ocular perfusion pressure (i.e., the difference between the blood pressure and the IOP). However, the occurrence of blood flow reversal in CRVO suggests that not only the ocular perfusion pressure comes into play for blood flow reversal, but also the inherent retinal vascular resistance. The retinal vascularization in the presence of glaucoma and/or elevated IOP has long been an active field of research. Some of the optical techniques that were used to conduct these studies were done with LSCI or LDF derived techniques which cannot determine the direction of blood flow, and therefore would wrongly assess the retinal vascularization in the presence of a flow reversal. Indeed, a diastolic backflow would be detected as a normally forward diastolic flow, so the average blood flow of an eye presenting arterial blood flow reversal as measured with an instrument insensitive to the direction blood flow, may remain in normal values over the cardiac cycle. Particularly, the resistivity index as measured with the Pourcelot ratio $(V_{\rm systole} - V_{\rm diastole})/V_{\rm systole}$ where $V$ is the velocity, would fail to account for the actual vascular resistance. A lower resistivity index would be found in the presence of diastolic backflow than in the absence of diastolic blood flow. On the other hand, we have found that the waveform profile in cases of backflow presents a distinctive characteristic. To this day, in our clinical study we have not observed in the absence of flow reversal such an early cancellation of blood flow that is followed by a significant rebound lasting until the end of diastole. Instead, in the absence of flow reversal, if there is a blood flow decrease to zero, blood flow would usually remain at zero until the end of diastole. Thus, there is reasonable hope that non-directional but high quality measurements of arterial blood flow could be sufficient to detect a retrograde flow.

We used the directional contrast to compare Doppler spectral traces in vessels with different orientations and concluded that the scattering regime of the dominant contribution to the Doppler spectrum is largely influenced by vessel geometry. The blood flow signal measured by LDH in in-plane vessels comes from multiply scattered light, but it seems that in out-of-plane vessels this signal originates from backscattered light. We reached this conclusion from the simple observation of the large asymmetry of energy distribution in spectrograms in areas with axially oriented flow. This probably implies that to derive the local velocity of blood flow, different scattering models should be used depending on the flow orientation. We hypothesize two reasons to explain why the contribution of multiply scattered light to the Doppler spectrum is lessened in axially oriented vessels. First, the effective light anisotropic factor in these blood vessels may be increased in consequence of the greater probed blood volume. A second possibility is that the preferred orientation of erythrocytes in flowing blood could change the light scattering model due to the anisotropic properties of erythrocytes~\cite{Fujii1999, Nilsson1998, Srinivasan2011, Cimalla2011}. A perspective opened up by the directional contrast is to realize blood velocity measurements in physical units with LDH in in-plane blood vessels. In out-of-plane vessels, Doppler measurements depend on backscattered light for which the Doppler angle must be known to retrieve the absolute velocity. However, the directional contrast can be used to identify in-plane sections of vessels where the Doppler signal originates only from multiply scattered light. The statistical properties of the Doppler spectrum of multiply scattered light should then be directly relatable to the absolute velocity of the local blood flow~\cite{MagnainCastelBoucneau2014}, which could be used to infer blood flow since the vessel diameter can be directly measured on LDH images.

In conclusion, we have presented a directional contrast that improves LDH angiographic images by providing qualitative information about the axial direction of local blood flow. This directional contrast relies on the strong Doppler spectral asymmetry of light backscattered in out-of-plane vessels and is helpful to identify the entrance and exit points of blood flow into the retina. Arterial blood flow reversals can be evidenced unambiguously, and quantitatively monitored with a high temporal resolution with spectrograms. We illustrated this ability in eyes affected by glaucoma and CRVO, and retinal vasculitis. Overall, LDH presents itself as a proficient alternative to Doppler ultrasound to study blood flow in the inner retinal vasculature.

\section*{Acknowledgments}
The authors thank Philippe Bonnin for helpful discussion. The Titan RTX graphics card used for this research was donated by the NVIDIA Corporation. This research was financially supported by the European Research Council (Synergy HELMHOLTZ \#610110), the Agence Nationale de la Recherche (FOReSIGHT ANR-18-IAHU-0001), and the Region Ile-de-France (EX047007 - SESAME 2019 - 4DEye).

\section*{Author contributions}
M.A. and M.P. initiated the research and secured funding. M.A. and L.P. worked on the optical setup. M.A. conducted the experiments. L.P. analyzed the data and wrote the article. All authors discussed the results and reviewed the manuscript.

\section*{Additional information}
\textbf{Competing interests} 
The authors declare no conflict of interest.

\section*{Supplementary Material}
\noindent
\textcolor{blue}{\href{https://youtu.be/H9IlPAJuOVA}{Supplementary Visualization 1}}. \newline
\textcolor{blue}{\href{https://youtu.be/5zs9gAzhUW8}{Supplementary Visualization 2}}. \newline

\bibliography{C:/Users/Allihies/Dropbox/Articles/DirectionalDopplerRetina/Biblio/Bibliography.bib}

\end{document}